%% file: example.tex
\documentclass[submission,copyright,creativecommons]{eptcs}
\input{header.tex}

\providecommand{\event}{FSFMA 2014} 
\usepackage{breakurl}             

\title{Verified Subtyping with Traits and Mixins}
\author{Asankhaya Sharma
\institute{Department of Computer Science}
\institute{National Univeristy of Singapore}
\email{asankhs@comp.nus.edu.sg}
}
\def\titlerunning{Verified Subtyping with Traits and Mixins}
\def\authorrunning{Asankhaya Sharma}
\begin{document}
\maketitle

\begin{abstract}
Traits allow decomposing programs into smaller parts and mixins
are a form of composition that resemble multiple inheritance.
Unfortunately, in the presence of traits, programming languages like
Scala give up on subtyping relation between objects.
In this paper, we present a method to check subtyping 
between objects based on entailment in separation logic.
We implement our method as a domain specific language in
Scala and apply it on the Scala standard library. We have
verified that 67\% of mixins used in the Scala standard library
do indeed conform to subtyping between the traits
that are used to build them.
\end{abstract}

\section{Introduction}
\input{intro.tex}

\section{Verified Subtyping}\label{formal}
\input{formal.tex}

\section{Implementation with SLEEK DSL}\label{implementation}
\input{sleek_interface.tex}

\section{Experiments}\label{experiments}
\input{subtype_exp.tex}

\section{Related Work}
\input{related.tex}

\section{Conclusions}
In this paper, we presented a method to enable verified subtyping in Scala.
Our method is based on
a reduction to entailment checking in separation logic.
We implemented a domain specific language (SLEEK DSL) in Scala
to enable programmers to check subtyping in their programs.
Using SLEEK DSL we carried out a study of the Scala standard library and verified
that 67\% of the mixins were composed of
traits that are in a subtyping relation.

\section*{Acknowledgements}
We thank Shengyi Wang for his 
prompt and useful feedback on the paper.
Cristian Gherghina 
and Chin Wei Ngan provided valuable comments
and suggestions on an early presentation of this work.

\bibliographystyle{eptcs}
\bibliography{scalasleek}
\end{document}

%% file: header.tex
\usepackage{ifthen}
\usepackage{times,frameit,epsfig,amsmath,amssymb,color,latexsym,graphics,wrapfig,clrscode3e}
\usepackage{frameit,amsmath,color,latexsym,graphics,wrapfig,bussproofs,proof,algorithm,algorithmic}
\usepackage{multirow}
\usepackage{hyperref}

\newcommand{\wnnay}[1]{}

\newcommand{\raznay}[1]{}

\newcommand{\chnay}[1]{}

\newcommand{\seppred}[2]{\ensuremath{#1\langle#2\rangle}}

\newcommand{\report}[1]{ }

\newcommand{\acm}[1]{ }

\newcommand{\hide}[1]{}
\newcommand{\hideie}[1]{}
\newcommand{\nil}{\btt{null}}

\newcommand{\emp}{\btt{emp}}
\newcommand{\veq}{\ensuremath{\equiv}}

\newcommand{\pure}{\ensuremath{\pi}}

\newcommand{\heap}{\ensuremath{\kappa}}

\newcommand{\constr}{\ensuremath{\Phi}}

\newcommand{\myit}[1]{\textit{#1}}

\def\sep{\code{*}}

\newcommand{\entrulen}[1]{[\underline{{\bf \scriptstyle ENT-#1}}]}

\newcommand{\hform}[3]{\ensuremath{{\mathrm{#1}}{::}{\btt{#2}}{\langle}{\mathrm{#3}}{\rangle}}}

\def\int{\code{int}}

\def\a{a}

\newcommand{\mynew}[1]{\code{new}~#1}

\newcommand{\code}[1]{{\small {\ensuremath{\tt #1}}}}

\newcommand{\btt}[1]{{\ensuremath{\tt #1}}}

\def\sep{\ensuremath{*}}

\def\inv{\myit{inv}}

\newcommand{\atom}{\alpha}

\def\myval{\mathrm{val}}
\def\mytrait{\mathrm{trait}}
\def\myextends{\mathrm{extends}}
\def\mysealed{\mathrm{sealed}}
\def\mywith{\mathrm{with}}
\def\mydef{\mathrm{def}}

\def\myprivate{\mathrm{private}}

\def\myvar{\mathrm{var}}
\def\myabstract{\mathrm{abstract}}
\def\myoverride{\mathrm{override}}

\def\mynew{\mathrm{new}}

%% file: intro.tex
Traits \cite{Ducasse06} have been recognized as a mechanism to support 
fine grained reuse in programming. Several programming languages (Scala, Fortress, Ruby, etc.) 
support the use of traits in some form or other. 
Traits and mixins provide support for code reuse
and composition that goes beyond classes and inheritance
in object oriented programs. In addition,
object oriented (OO) programs themselves are notoriously hard to
verify in a modular fashion. 
Recently \cite{ChinOO,Parkinson08:POPL,Distefano08:OOPSLA} separation logic based approach has yielded success
in verification of object oriented programs. This include support for verifying inheritance 
and behavior subtyping, in conformance with OO paradigm. 
In this paper, we extend the work done on
verification of 
OO programs in separation logic to verify subtyping with traits
and mixins.

Below we consider an example that illustrates the problem of subtyping with traits
and mixins.
The $\mathit{ICell}$
trait captures an object with an integer value that can be accessed with $get$ and $set$ methods.
The $BICell$ trait provides a basic implementation for $ICell$, while the $Double$ and $Inc$ traits
extend the $ICell$ trait by doubling and incrementing the integer value respectively.
\[
\begin{array}{l}
\mytrait ~~ICell~~\{ \\ 
~\mydef ~~get(): Int \\
~\mydef ~~set(x: Int) \} 
\end{array}
\qquad
\begin{array}{l}
\mytrait ~~BICell ~~\myextends ~~ICell ~\{ \\ 
~\myprivate ~~\myvar ~~x: Int = 0 \\ 
~\mydef ~~get() \\
~~~~~ \{~x~\}\\
~~\mydef ~~set(x: Int) \\
~~~~~\{ ~this.x = x ~\}\\
\}
\end{array}
\qquad
\begin{array}{l}
\mytrait ~Double~ \myextends ~ICell ~\{\\
~\myabstract ~\myoverride ~\mydef ~set(x: Int)\\ 
~~~~~~~\{ ~\mathrm{super}.set(2 * x) \}\\
~~\}\\
\mytrait ~Inc ~\myextends~ ICell~ \{\\
~\myabstract ~\myoverride ~\mydef ~set(x: Int)\\
~~~~~~~\{ \mathrm{super}.set(x + 1) \}\\
~~\}
\end{array}
\]

These traits are used in the following class mixins. The integer value field of the objects of $OddICell$ mixin
is always odd, while the value is even for objects of $EvenICell$ mixin.
\[
\begin{array}{l}
 \mathsf{class} ~~OddICell ~\mathsf{extends} ~~BICell ~\mathsf{with} ~~Inc ~\mathsf{with}
 ~~Double\\
 \mathsf{class} ~~EvenICell ~\mathsf{extends} ~~BICell ~\mathsf{with} ~~Double ~\mathsf{with}
 ~~Inc\\
 \end{array}
\]

In the presence of traits, the type system of Scala is not strong enough to distinguish between
accepted uses of the traits. This can be illustrated by the following example.
\vspace{-2mm}
 \[
 \begin{array}{l}
  \mydef ~m ~(c: ~BICell ~\mywith ~Inc ~\mywith ~Double): ~Int = \{c.get \}\\ 
  \myval ~oic = \mynew ~OddICell\\ 
  \myval ~eic = \mynew ~EvenICell \\
  m(oic) ~~//~ Valid \\
  m(eic) ~~//~ Valid\\
 \end{array}
\] 

The method $m$ can be called with an object of both mixins $EvenICell$ and $OddICell$,
even though the expected object (c) type is a supertype of $OddICell$ and not $EvenICell$. 
Thus, the type system in Scala cannot distinguish between the two calls made to method $m$
as it does not check for subtyping between the objects. The key contribution of this paper
is to present a method for checking subtyping in the presence of traits and mixins in Scala.
In 
section \ref{formal}, we present an approach based on entailment in separation logic
to verify subtyping. In section \ref{implementation}, we present a domain specific
language which is embedded in Scala and can support verified subtyping with traits and mixins.
We apply our technique to the mixin class hierarchies in the Scala standard library and verify 
subtyping in 67\% of the traits as shown in section \ref{experiments}. Our complete development 
including the source code of the domain specific language and all examples are available on-line at the following URL.

\centerline{ \url{http://loris-7.ddns.comp.nus.edu.sg/\~project/SLEEKDSL/}}

%% file: formal.tex
We consider a core language based on \cite{ChinOO} for formalizing our approach.
As shown in figure \ref{fig.formal}, to simplify the presentation we focus only on type information for traits
and mixins while ignoring all other features in our core language. 
We also assume that all classes are part of mixin compositions and only traits are used to create mixins.
Since, existing approaches \cite{ChinOO} can handle class based single inheritance, we focus only on mixin compositions in this paper. The rest of the constructs in the core language are related to 
predicates ($\constr$) in separation logic. Each trait (and mixin) $C$ can be represented by
a corresponding predicate $\seppred{\myit{C}}{v^*}$.
\vspace{-4mm}
\begin{figure}[thb]
\begin{center}
\begin{minipage}{0.6\textwidth}
\begin{frameit}
\vspace{-4mm}
 \[
 \begin{array}{ll}
mixin &::= ~class~C~[extends~C_1]~[with~C_2]^* \\
pred &::=   \seppred{\myit{C}}{v^*}~{\veq}~ 
 \constr ~~  [\code{\inv} ~~\pure] \\
 \constr &::= \bigvee~(\exists w^*{\cdot}\heap{\wedge}\pure)^* \\
\heap &::= \emp ~|~ \seppred{C}{v^*}
~|~ \heap_1 \sep \heap_2 
\\
\pure &::= \atom ~|~ \pure_1{\wedge}\pure_2 \quad \atom ::= \beta ~|~ \neg \beta\\
\beta &::= 
v_1 {=} v_2 ~|~ v{=}\nil ~|~ a{\leq}0 ~|~ a{=}0  
\\
\a &::=  k \mid k{\times}v \mid \a_1+\a_2 \hide{\mid -a}

\\
\end{array}
\]
\vspace{-4mm}
\caption{Core Language for Traits and Mixins}
\label{fig.formal}
\end{frameit}
\end{minipage}
\end{center}
\vspace{-4mm}
\end{figure}

Predicates based on separation logic are sufficient to specify mixins
because of class linearization in Scala \cite{Martin:Scala}.
After class linearization a mixin class composition (unlike multiple inheritance) has a single linear hierarchy. 
In the case of our running example, the mixins  give rise to the following linearizations:
 
 \[ 
\begin{array}{l}
OddICell \leftarrow Double \leftarrow Inc \leftarrow BICell\\ 
\seppred{\myit{OddICell}}{this} \equiv \seppred{\myit{BICell}}{this,v} \sep \seppred{\myit{Inc}}{v,v_1} \sep \seppred{\myit{Double}}{v_1,\nil} \\
EvenICell \leftarrow Inc \leftarrow Double \leftarrow BICell\\
\seppred{\myit{EvenICell}}{this} \equiv \seppred{\myit{BICell}}{this,v} \sep \seppred{\myit{Double}}{v,v_1} \sep \seppred{\myit{Inc}}{v_1,\nil} \\
\end{array}
\]

 A mixin class composition can be treated as a single inheritance hierarchy  based on the
 linearization and thus, subtyping between the mixins can be decided by checking 
 the entailment based on separation logic predicates. 
 In case of our running example, the call to method $m$ 
 is valid with $oic$ object but not the $eic$ object as the following entailments show. 
\[ 
\begin{array}{l}
\seppred{\myit{OddICell}}{oic} \vdash
\seppred{\myit{BICell}}{c,v} \sep \seppred{\myit{Inc}}{v,v_1} \sep \seppred{\myit{Double}}{v_1,\nil}
\quad Valid
\\
\seppred{\myit{EvenICell}}{eic} \vdash
\seppred{\myit{BICell}}{c,v} \sep \seppred{\myit{Inc}}{v,v_1} \sep \seppred{\myit{Double}}{v_1,\nil}
\quad Invalid
\end{array}
\]
We now show how the problem of checking subtyping between objects belonging to two different
mixins is reduced to an entailment between the corresponding predicates in separation logic.
This entailment can be checked with the help of existing solvers for separation logic
(like SLEEK \cite{Chin11}). The entailment rule for checking subtyping with traits and mixins is given
in figure \ref{fig.subtype}. An object of mixin $C$ is a subtype of mixin $D$ when 
the entailment between their corresponding predicates in separation logic is valid.

\begin{figure}[thb]
\begin{center}
\begin{minipage}{0.6\textwidth}
\begin{frameit}
\vspace{-4mm}
 \[
\frac{\begin{array}{c}
\entrulen{Subtype-Check}\\
 ~class~C~[extends~C_1]~[with~C_2]^* \\
 ~class~D~[extends~D_1]~[with~D_2]^* \\
{\seppred{\myit{C}_1}{this,v_1}[\sep \seppred{\myit{C}_2}{v_1,v_2}]^* \vdash
\seppred{\myit{D}_1}{this,u_1}[\sep \seppred{\myit{D}_2}{u_1,u_2}]^*}
\end{array}}
{C <: D}
\]
\vspace{-4mm}
\caption{Checking Subtyping with Entailment}
\label{fig.subtype}
\end{frameit}
\end{minipage}
\end{center}
\vspace{-4mm}
\end{figure}

Entailment checking in separation logic can be used to decide subtyping with traits and mixins. But in order
to integrate subtyping support inside Scala we face some engineering challenges. 
In particular, it is
too restrictive and infeasible to do this kind of checking for all the mixins. 
This requires support for
selective subtyping as all mixins will not satisfy the subtype relation.
In order to provide the programmer the choice of checking subtyping usage 
in their methods we have implemented an embedded domain specific language (DSL) in Scala.
This DSL uses the SLEEK entailment checker for checking the validity of entailments in separation logic.
In the next section we describe the SLEEK DSL and how it is integrated in Scala.

%% file: sleek_interface.tex
\begin{figure}
\begin{center}
\includegraphics{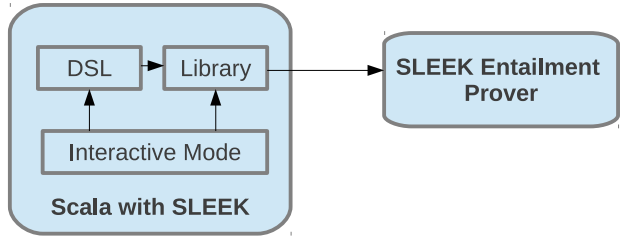}
\end{center}
\caption{Overview of SLEEK DSL}
\label{fig.dia}
\end{figure}

Our proposal is based on embedding a domain specific language (SLEEK DSL) in Scala.
As shown in figure \ref{fig.dia}, a Scala library (SLEEK lib) interfaces directly
with the external application - the SLEEK entailment prover. In addition, we extend
Scala with a DSL (SLEEK DSL) which makes use of the Scala library
to provide the entailment checking feature inside Scala programs.
Further, for using with the Scala interpreter we provide an interactive mode (SLEEK inter)
which uses the SLEEK DSL and library to enable interactive entailment proving.
Thus, the implementation of the verified subtyping in Scala with SLEEK
has three
main components:
\begin{itemize}
  \item a Scala library that supports all SLEEK interactions
  \hide{including the SLEEK interactive mode}
  \item a domain specific language (DSL) implemented in Scala that models the SLEEK input language.
  With this DSL we get for free embedded type checking in Scala.
  \item a helper library designed for the Scala interpreter. The library runs SLEEK in
  interactive mode in the background to provide seamless integration with Scala.
\end{itemize}

In short, the SLEEK library provides basic functionality for constructing Scala objects representing
separation logic formulas. 
The entailment checking method is in fact the actual
front-end for SLEEK. It takes two Scala objects representing separation logic formulas,  
translates them to the SLEEK input language and invokes SLEEK. The result and the possible residue
is captured and parsed using the Scala parser combinator library to extract the Scala representation.
To facilitate a better syntax for writing formulas and support for richer interplay with the Scala
types we present a domain specific language, SLEEK DSL implemented on top of
the Scala library.
We will outline the SLEEK DSL by
presenting how an entailment checking can be encoded in our DSL.

\subsection{SLEEK DSL}
\label{sleek.dsl}

As an example consider the following entailment check between two separation logic
formulas defined using SLEEK DSL.
\[
\myval ~r = ~~~~\hform{x}{node}{\_,null} \vdash \hform{x}{ll}{m} ~\mathrm{\&\&}~ \mathrm{m} {===} 1 
\]

It encodes an entailment between two formulas, one describing a single heap node, an instance of a
data structure called $\mathit{node}$. The second formula describes a state in which $\mathrm{x}$ is the
root pointer of for a data structure described by the $\mathit{ll}$ predicate. This predicate abstracts a
linked list of size $\mathrm{m}$.

SLEEK DSL relies on the functions defined in the SLEEK Library to create new easy to use
operators that provide a more user friendly syntax. A special operator, the double colon ($::$) is
used to describe the points-to relation commonly used for heap nodes.  It also provides the usual
arithmetic (e.g. $+$,$-$) and boolean (e.g. $\&\&$, $||$, $===$, $!==$, $\vdash$)  operators to
help in constructing the separation logic formula. 
The notation used in the DSL is similar to the one used for SLEEK in \cite{Chin11}. The use of a
 DSL allows easy intermixing of SLEEK formulas with other Scala types. We use implicit conversions
 between types (e.g. from $\mathit{scala.Int}$ to $\mathit{formula[IntSort]}$) to make it even
 easier to use these formulas in Scala programs.

\hide{
The value $\mathit{r}$ encodes an entailment check between two separation logic formulas
naturally specified in the SLEEK DSL.$\mathit{r}$ is of type formula and has a property
$\mathit{isValid}$ defined for it.}  Furthermore, our library provides a definition for the
$\mathit{isValid}$ method in the formula class. In order to check the  validity of the above
entailment it is sufficient to call $\mathit{r.isValid}$ which feeds the entailment to
SLEEK and converts the result back into a $\mathit{scala.Boolean}$ for use as a conditional. Implicit
methods provide an easy mechanism to convert from one type of object to the desired type. This
enables the support for a SLEEK like syntax within Scala.
Formulas allow for a variety of types for the parameters used (such
as $\mathit{x}$ and $\mathit{m}$). In the Scala library these types are grouped under the following
type hierarchy.
\[
\begin{array}{l}
\mysealed ~~\mytrait~~ Top \\
\mytrait ~~BoolSort ~~\myextends ~~Top \\ 
\mytrait ~~IntSort ~~\myextends ~~Top \\
\mytrait ~~BagSort ~~\myextends ~~Top \\
\mytrait ~~ShapeSort ~~\myextends ~~Top \\
\mytrait ~~Bottom ~~\myextends ~~BoolSort \\
~~~~~~~~~~\mywith ~~IntSort ~~\mywith ~~BagSort ~~\mywith ~~ShapeSort 
\end{array}
\]
This trait allows the embedding of the types used in the separation logic formula as Scala types.
By defining the various operators using these types, soft type checking for SLEEK
formulas is automatically ensured by the underlying Scala type system. 
The benefit of using a DSL is that it provides a simpler syntax and familiar look and feel for
the user of the library. The formula represented by the DSL is also much more concise. 

The SLEEK DSL allows programmers to verify entailments written in separation logic.
In addition, programmers can use 
the DSL to encode subtyping check as an entailment check in separation logic
as described in section \ref{formal}.

\hide{As an example consider the definition of the $\mathit{+}$ operator.
$$
\mydef ~+~ (that: \mathrm{typedformula}[IntSort]):\mathrm{typedformula}[IntSort] 
$$
This operator takes two formulas, each of type $\mathit{IntSort}$, and gives a resulting formula
which is also of type $\mathit{IntSort}$. Note that overloaded operators are defined as class
methods, therefore one of the operands is the object pointed to by $\mathit{this}$ and the second is
the method argument. Similarly a Boolean operator can be defined from any type within the hierarchy
say $\mathit{A}$ to $\mathit{BoolSort}$. Once we have a formula representing an entailment
we can check the validity of the formula using the $\mathit{isValid}$ method. 
The property
$\mathit{isValid}$ is implemented using SLEEK Library. It calls SLEEK to check the entailment
and parse the result returned by SLEEK. The result is converted into a Boolean value so that it can be used
in Scala . 
The
 typed formula is converted to the untyped notation used in the SLEEK library and then passed to the
$\mathit{checkvalid}$ method.
}

\hide{However, we are still using SLEEK in a non-interactive mode.} 

\subsection{SLEEK Interactive Mode}
\label{sleek.inter}
The Scala runtime provides a good interpreter for rapid prototyping which can be used from the
command line. Similarly, SLEEK also has an interactive mode in which it accepts commands and gives
the results back to command line. 
In order to make SLEEK's interactive mode available to the Scala interpreter, we provide a helper
library that hides the extra intricacies incurred by using SLEEK interactively. 
The benefit of using the interactive mode is that the user defined predicates 
and data types will not be defined again with each call to $\mathit{isValid}$ method. This makes the interactive mode 
of SLEEK DSL faster when compared to calling the same function from the basic SLEEK library.
\hide{
We believe that the three components of the library (Base, DSL, and Interactive Mode) cater for
most usecases and allow for possible extensions. For example, it is possible to construct libraries
written in the SLEEK DSL that contain definitions for most of the common data structures and
predicates (e.g. Linked List, Queue, Doubly Linked List, Binary Search Tree etc.). Such definitions
could then be directly imported, minimizing further on the overhead of using this library.}

Our implementation for verified subtyping integrates into Scala
as an API (SLEEK library), as a language (SLEEK DSL)
and as an interpreter (SLEEK Interactive mode).  This provides programmers
the ability to use our procedure in different ways as desired.

%% file: subtype_exp.tex
We have used SLEEK DSL to verify subtyping of mixin compositions from the Scala standard library. 
To the best of our knowledge this it the first such study of subtyping in Scala.
The following table presents the results. The first column is the name of the class hierarchy.
The second column lists the total number of mixins in the hierarchy, while the third column gives
the number of mixins for which we can verify that the subtyping relation holds.
The last column gives the percentage of mixins with subtyping. 
\vspace{-4mm}
\[
\begin{array}[t]{|c|c|c|c|}
\hline
\textit{Class Hierarchy} & \textit{Total Num of Mixins} & \textit{Mixins with Subtyping} & \textit{Percentage}\\
\hline
\textit{Exceptions} & 11 & 11 & 100\\
\textit{Maths} & 5 & 4 & 80 \\
\textit{Parser Combinator} & 6 & 6 & 100 \\
\textit{Collections} & 27 & 12 & 44  \\
\hline
\textit{Total} & 49 & 33 & 67  \\
\hline
\end{array}
\]
As an example of mixin hierarchy whose subtyping relations are verified consider the following
which represents the maths library in Scala. The only mixin which breaks the subtyping relation is  
\text{PartialOrdering}. Rest of the mixins can be verified to respect the expected subtyping. Thus 
we have verified that subtyping holds for 4 out of 5 mixins that are part of math class hierarchy. 
\vspace{-2mm}
\[
\begin{array}[t]{l}
\text{Equiv is SUPERTYPE of PartialOrdering}\\
	\qquad \text{PartialOrdering is NOT SUPERTYPE of Ordering}\\
		\qquad \qquad  \text{Ordering is SUPERTYPE of Numeric}\\
			\qquad \qquad \qquad  \text{Numeric is SUPERTYPE of Integral}\\
			 \qquad \qquad \qquad \text{Numeric is SUPERTYPE of Fractional}
\end{array}\vspace{-2mm}
\]

%% file: related.tex
The work that comes closest to our method for checking subtyping is the work of Bierman et.al
\cite{bierman2010semantic}, they provide a mechanism to use SMT solvers for deciding
subtyping in a first order functional language.
On the other hand, we use SLEEK an entailment checker for separation logic
to decide subtyping between traits and mixins. SMT solvers
have also been used \cite{backes2011automatically} for verifying typing constraints.
Similar to our implementation of SLEEK DSL, the $Scala^{Z3}$ proposal of K\"{o}ksal et. al \cite{Koksal} integrates the
Z3 SMT solver into Scala. Although the integration is similar, the two solvers have different
focuses: Z3 is a general SMT solver, while SLEEK is a prover for separation logic. 
\hide{
In this work we have used the implicit arguments allowed by Scala to showcase possible applications
of the SLEEK integration. We have proposed a lightweight approach to verifying methods with
implicit arguments. Oliviera et. al \cite{oliveira} do an in-depth analysis of implicits as a
generic programming mechanism. Through the implicit calculus they provide  a powerful type-system
for generic programming with implicits in a higher order setting. However Scala does not use rule
abstractions for scoping of implicits, in Scala, implicit arguments can only be used in definitions.
We handle implicit arguments in methods using the existing inference mechanism present in SLEEK.}

Another line of work is on specification and verification of traits and mixins.
Damiani et. al explore trait verification in \cite{Damiani}. They observe the need for multiple
specifications and introduce the concept of proof outline. They support a trait based language with
limited composition - symmetric sum of traits and trait alteration. 
Our work does not directly address the issue of trait verification but checking subtyping
is essential part of OO verification using separation logic \cite{ChinOO}.
We believe that dynamic specifications of \cite{ChinOO} along with verified subtyping can be used
to verify traits and mixins.
Behavior subtyping is a stronger notion of subtyping between objects.
The approach of lazy behavioral subtyping \cite{Dovland:2011} can support incremental
verification of classes in presence of multiple inheritance. However, this is overly restrictive for
mixin compositions in Scala and our method provides a more
flexible support for subtyping in Scala.
